\documentclass[amsmath, amssymb, preprintnumbers, showpacs, showkeys, aps,prb,superscriptaddress,twocolumn]{revtex4-1}
\usepackage[utf8]{inputenc}
\usepackage{listings}
\usepackage{footmisc}
\usepackage{enumerate}
\usepackage{latexsym}
\usepackage{braket}
\usepackage{graphicx}
\usepackage[caption=false]{subfig}
\usepackage[colorlinks=True,linkcolor=red,citecolor=blue,urlcolor=blue]{hyperref}
\usepackage{blkarray}
\usepackage{array}
\usepackage{color}
\usepackage[normalem]{ulem}
\usepackage{tikz} 
\usepackage{amsthm} 
\theoremstyle{definition} 

\newcommand{\ti}[1]{\textsubscript{#1}}

\newcommand{\f}[1]{\boldsymbol{#1}}
\newcommand{\nk}[1]{\mathrm{#1}}
\newcommand{\rf}{\boldsymbol{r}}
\newcommand{\kf}{\boldsymbol{k}}

\newcommand*\diff{\mathop{}\!\mathrm{d}}

\newcommand{\specialcell}[2][c]{%
  \begin{tabular}[#1]{@{}c@{}}#2\end{tabular}}

\newtheorem{defi}{\textit{Definition}} 

\date{\today}

\begin{document}

\title{Elementary band representations for the single-particle Green's function of interacting topological insulators}

\author{Dominik Lessnich}
\email{lessnich@itp.uni-frankfurt.de}
\affiliation{Institute of Theoretical Physics, Goethe University Frankfurt, Max-von-Laue-Strasse 1, 60438 Frankfurt am Main, Germany}

\author{Stephen M. Winter}
\email{winters@wfu.edu}
\affiliation{Institute of Theoretical Physics, Goethe University Frankfurt, Max-von-Laue-Strasse 1, 60438 Frankfurt am Main, Germany}
\affiliation{Department of Physics, Wake Forest University, 1834 Wake Forest Road Winston-Salem, North Carolina 27109-7507, USA}

\author{Mikel Iraola}
\affiliation{Donostia International Physics Center, 20018 Donostia-San Sebastian, Spain}
\affiliation{Department of Physics, University of the Basque Country UPV/EHU, Apartado 644, 48080 Bilbao, Spain}

\author{Maia G. Vergniory}
\affiliation{Donostia International Physics Center, 20018 Donostia-San Sebastian, Spain}
\affiliation{IKERBASQUE, Basque Foundation for Science, Maria Diaz de Haro 3, 48013 Bilbao, Spain}

\author{Roser Valentí}
\email{valenti@itp.uni-frankfurt.de}
\affiliation{Institute of Theoretical Physics, Goethe University Frankfurt, Max-von-Laue-Strasse 1, 60438 Frankfurt am Main, Germany}

\begin{abstract}

We discuss the applicability of elementary band representations (EBRs) to diagnose spatial- and time-reversal-symmetry protected topological phases in interacting insulators in terms of their single-particle Green's functions. We do so by considering an auxiliary non-interacting system $H_{\textrm{T}}(\mathbf{k}) = -G^{-1}(0, \mathbf{k})$, known as the topological Hamiltonian, whose bands can be labeled by EBRs. This labeling is robust if neither (i) the gap in the spectral function at zero frequency closes, (ii) the Green's function has a zero at zero frequency or (iii) the Green's function breaks a protecting symmetry. We demonstrate the use of EBRs applied to the Green's function on the one-dimensional Su-Schrieffer-Heeger model with Hubbard interactions, which we solve by exact diagonalization for a finite number of unit cells. Finally, the use of EBRs for the Green's function to diagnose so-called symmetry protected topological phases is discussed, but remains an open question.

\end{abstract}

\maketitle

\section{Introduction} \label{intro}

Non-interacting topological insulators are well understood in terms of band topology.~\cite{ryu2010topological,Rev_scheng,hasan_rev,Moore_rev, kane_rev} 
An insulator is called topologically trivial if it is possible to continuously deform its band structure and corresponding eigenstates to those of an atomic insulator without closing the energy gap or breaking a symmetry. On the other hand, a phase characterized by a non-trivial topological invariant indicates an obstruction to such a continuous deformation to an atomic insulator. The Chern number, first proposed for the integer quantum Hall effect, is the most common example for a topological invariant.~\cite{Thouless_QHE}
Actually, all non-interacting topological phases protected by combinations of time-reversal symmetry (TR), particle-hole symmetry (PH) and chiral symmetry (CS), i.e. which belong to one of the ten Cartan-Altland-Zirnbauer (CAZ) symmetry classes,~\cite{caz} have been classified by K-theory~\cite{CAZ_zirnbauer,Kitaev_ktheory}.

An important class of topological insulators are topological crystalline insulators which
are protected by spatial symmetries.~\cite{fu_tci} 
In three dimensions non-interacting topological crystalline insulators have been systematically investigated in all 230 space groups, with and without TR present, in the formalism of topological quantum chemistry (TQC) \cite{Bradlyn2017, EBRs} in terms of elementary band representations (EBRs) \cite{Zak1980,zak1999,zak2001} or equivalently in the formalism of symmetry indicators~\cite{ashvin_sym,ashvinsurface} or the algorithm in Ref.~\onlinecite{slager2017}.
These formalisms are based on the fact that Bloch wave functions at high symmetry k-points can be classified by irreducible representations (irreps) of the little group of these k-points.
In this way spatial symmetries place constraints on the connectivity of the bands in the Brillouin zone, which can be used to identify those band structures that are compatible with an atomic insulator.

The applicability of such approaches when interactions are included is, however, unclear.  
In principle, non-interacting insulators have a very simple structure. Their ground state wave function is given by a Slater determinant of all single-particle states below the Fermi level. Thus, the gap in the single-particle spectrum makes the ground state wave function unique. 
To decide if two non-interacting insulators are topologically equivalent is the same as investigating whether the corresponding Hamiltonians can be smoothly connected while maintaining symmetries and maintaining the gap. The gap manifests in terms of (i) the many-body ground state staying gapped, or (ii) the presence of a gap in the single-particle excitations. Both features are equivalent in the absence of interactions. Hence it is sufficient to analyze the topological properties of the map $\kf \mapsto \mathcal{H}_0(\kf )$, where $\kf$ is a reciprocal wavevector and $\mathcal{H}_0(\kf )$ is the Bloch Hamiltonian. For non-interacting systems, this corresponds to investigating the topological properties of the single-particle Matsubara Green's function $G(i\omega, \kf)$, which for the non-interacting case is given as:
 
\begin{equation}
G_0(i \omega, \kf ) = \big( i \omega - \mathcal{H}_0(\kf ) \big)^{-1}
\label{G0def}
\end{equation}
where $i \omega$ denotes the Matsubara frequency. 
In the presence of interactions investigating the adiabatic connectivity of Hamiltonians while the ground state stays gapped and investigating the Green's function is a priori not equivalent anymore.

In a more general context, the concept of symmetry protected topological (SPT) phases~\cite{wen2009,pollman2010,kitaev1D,Kitaev1D_p2,turner_1D,wen2011,sentil2015} has been introduced to investigate the smooth connectivity of gapped, short range entangled phases while maintaining symmetries.
Alternatively, the topological characterization of the full interacting single-particle Matsubara Green's function $G(i\omega, \kf)$
was put forward in Refs.~\onlinecite{Volovik2003,Volovik2009He,Volovik2010,gurarieG,wang10,wang12inv,wang12,Wang_2013}. For the CAZ symmetry classes it was shown that one obtains the same topological classification for the Green's function as for non-interacting Hamiltonians i.e. $\mathbb{Z}$, $\mathbb{Z}_2$ or $0$.~\cite{gurarieG} 
A similar Green's functions-based framework for identifying spatial-symmetry-protected topological phases in interacting systems has, however, not been fully explored. 

For this purpose, in the present work we first demonstrate that an EBR classification --successfully implemented to diagnose band topology of non-interacting topological insulators in the framework of TQC--, can be applied to $G(i \omega,\kf)$ and, secondly, we  discuss its suitability/practicability to identify spatial- and time-reversal-symmetry protected topological insulating phases.

The paper is organized as follows. In Sec.~\ref{interactingG} we establish the conditions that
$G(i \omega,\kf)$ needs to fulfill to define topological invariants in terms of $G(i \omega,\kf)$ and, we review the concept of a topological Hamiltonian. This analysis sets the framework for  the EBR classification of Green's functions.
In Sec.~\ref{G_symprop} we investigate the implications that the spatial symmetries of the many-body Hamiltonian have on $G(i \omega,\kf)$. In Sec.~\ref{hteigv} we discuss the EBR-based classification of the topological Hamiltonian and argue about its use and limitations. In Sec.~\ref{demo} we analyze the interacting Green's function of the \hbox{one-dimensional} Su-Schrieffer-Heeger model~\cite{ssh_model} with Hubbard interactions (SSH+U) within the framework of TQC, diagnosing its topological phases
by making use of the spatial inversion symmetry present in the model and, in Sec.~\ref{Conclusions} we present our conclusions.

\section{Interacting Green's function and topological Hamiltonian}
\label{interactingG}

For the analysis of symmetry-protected topological invariants in terms of Green's functions~\cite{Volovik2003,Volovik2009He, Volovik2010,gurarieG,wang10,wang12inv,wang12,Wang_2013} it is assumed that the exact ground state of the many-body Hamiltonian is unique and the chemical potential is included in the many-body Hamiltonian.
Further, we consider the zero-temperature limit so the discrete Matsubara frequencies $i \omega_n$ become continuous $i \omega_n \to i \omega$. 
In this case, the topological invariants are well defined and maintained under continuous changes of the Green's function as long as the GNSC-conditions (gapped, non-singular, symmetries preserved, continuously differentiable) defined below, are fulfilled. 
\begin{defi}\mbox{}\label{GNSCc}
A Matsubara Green's function $G(i \omega, \kf) $ (matrix) with associated spectral function $A(\omega, \kf )$ together with a set of protecting symmetries fulfills the GNSC-conditions if all following conditions hold: 
\begin{enumerate}\label{Gconds}
    \item There is a non-zero gap in $A(\omega, \kf )$ at zero frequency, i.e. there exists an $\epsilon > 0$ such that $A(\omega, \kf)=0$ for all \hbox{$\omega \in \left[ - \epsilon , \epsilon \right]$} for all $\kf$.
    \item $G(0,\kf)$ is non-singular which implies that all eigenvalues are non-zero for all $\kf$.
    \item $G$ does not break a symmetry contained in the set of protecting symmetries.
    \item $G(i \omega, \kf) $ is continuously differentiable in $\kf $ for all $i \omega$.
\end{enumerate}
\end{defi}
It may be noted that G does not break a spatial symmetry, meaning that $G(i \omega, \kf)$ commutes with the band representation (or quasiband representation) matrices as discussed in Sec.~\ref{G_symprop}. For the implications of TR, PH and CS on $G$, see Ref.~\onlinecite{gurarieG}.
For completeness, we review in Appendix~\ref{anaG} analytic properties of the Green's function.

In Ref.~\onlinecite{wang12} it was shown that it is sufficient to focus on the Green's function at zero frequency to obtain the topological invariants for the CAZ symmetry classes.
Equivalently it is possible to define an auxiliary non-interacting Hamiltonian~\cite{Wang_2013}  -- the topological Hamiltonian -- which has the same topological invariants as the full interacting single-particle Green's function
\begin{equation}
H_{\textrm{T}}(\mathbf{k}) = -G^{-1}(0, \mathbf{k}).
\label{htopdef}
\end{equation}
If the spectral function is gapped, $H_{\textrm{T}}(\mathbf{k})$ is a Hermitian matrix with absolute values of the eigenvalues bounded from below for all k-points (see Appendix~\ref{anaG}).
 
 The analysis of topological invariants in terms of the Green's function has been applied to a variety of model systems in which the topology is protected by symmetries in the CAZ symmetry classes \cite{gurarie1d,prlsshpu,limitG1,mertz19,gex1,gex2,gex3,gex4,gex5,gex6} and it has been shown that, if at least one of the GNSC-conditions is violated, the respective CAZ invariant is not well defined anymore.~\cite{gurarieG,gurarie1d}
 An alternative approach based on the local in-gap Green's function has been developed in Ref.~\onlinecite{slager2015}.
 In what follows, we analyze $G(i\omega,\kf)$ through the topological Hamiltonian $H_T$ within the framework of TQC and investigate the range of applicability of the method.

\section{Spatial symmetries of the Matsubara Green's function}
\label{G_symprop}

In this section, we recall the action of spatial symmetries on the Green's function and $H_{\nk{T}}(\kf)$. We show that these always transform in the same way as non-interacting Bloch Hamiltonians. For clarification of our notation see Appendix~\ref{anaG}.
For a given space group $\mathfrak{G}$, the spatial symmetries $h =\{ R|\mathbf{v} \} \in \mathfrak{G}$ act in real space as $\rf \rightarrow R \rf + \mathbf{v}$. If we associate a unitary operator $U_h$ with the symmetry operation $h \in \mathfrak{G}$, then $U_h$ acts on creation and annihilation operators as:
\begin{align}
U_h c^{\dagger}_{\kf \alpha} U_h^{\dagger} &= \sum_{\beta} \rho_{\mathfrak{G}}^{\kf}(h)_{\beta \alpha} c^{\dagger}_{\kf' \beta}, \\
U_h c_{\kf \alpha} U_h^{\dagger} &=  \sum_{\beta} \big(\rho_{\mathfrak{G}}^{\kf}(h)^* \big)_{\alpha \beta} c_{\kf' \beta},
\end{align}
where $(^*)$ denotes the conjugate transpose of a matrix, and $\alpha$ labels e.g.~band and spin indices.
Each transforms according to the band representations \hbox{ $\rho_{\mathfrak{G}}^{\kf}(h)$. \cite{Bradlyn2017,EBRs}}
Note that for simplicity we consider the Bloch-like wave function associated with $c^{\dagger}_{\kf \alpha}$ and $c_{\kf \alpha}$ to originate from exponentially localized, symmetry consistent Wannier functions (see Appendix~\ref{anaG}). 
In principle, the discussion is also applicable to isolated sets of topological bands without an atomic limit, i.e. bands that transform as representations that can not be written as linear combinations of EBRs with positive integer coefficients. In this case the topological set of bands correspond to a so-called quasiband representation, i.e. any solution of the compatibility relations.~\cite{Bradlyn2017,EBRs}

Here and in the following we set $\kf' \equiv R \kf$, with $\kf'$ being the corresponding k-point in the first Brillouin zone. The representations $ \rho_{\mathfrak{G}}^{\kf}(h)$ are sets of unitary matrices with the dimension given by the number of orbitals in the unit cell (including spin). See Ref.~\onlinecite{EBRs} for the explicit construction of the band representation matrices. 
A general Hamilton operator $H$ is invariant under the action of $h$ if it commutes with $U_h$.

Let us first review the consequences of unitary spatial symmetries on non-interacting Hamiltonians. These can be written as
\begin{equation}
H_0 = \sum_{\kf} H_0^{\kf} = \sum_{\kf} \sum_{\alpha \beta}  c^{\dagger}_{\kf \alpha} \big( \mathcal{H}_{0}(\kf) \big)_{\alpha \beta} c_{\kf \beta}.
\end{equation}
where $\mathcal{H}_0(\kf)$ is the Bloch Hamiltonian matrix. The Hamiltonian transforms under symmetries as
\begin{align}
H_0 &=  U_h H_{0} U_h^{\dagger} \nonumber \\
&= \sum_{\kf \alpha \beta} U_h c^{\dagger}_{\kf \alpha}  \big( \mathcal{H}_0(\kf) \big)_{\alpha \beta} c_{\kf \beta} U_h^{\dagger} \nonumber \\
&= \sum_{\kf \alpha \beta \gamma \delta}  c^{\dagger}_{\kf' \gamma} \rho_{\mathfrak{G}}^{\kf}(h)_{\gamma \alpha} \big( \mathcal{H}_0(\kf) \big)_{\alpha \beta}  \rho_{\mathfrak{G}}^{\kf}(h)^*_{\beta \delta} c_{\kf' \delta} ,
\end{align}
So for the Bloch Hamiltonian must transform with the band representations as
\begin{equation}
\big( \mathcal{H}_0(\kf^\prime) \big)_{\gamma \delta} = \sum_{\alpha \beta } \rho_{\mathfrak{G}}^{\kf}(h)_{\gamma \alpha} \big( \mathcal{H}_0(\kf) \big)_{\alpha \beta} \rho_{\mathfrak{G}}^{\kf}(h)^*_{\beta \delta}.
\label{h0br}
\end{equation}
To determine the irreps of wave functions in a particular k-point $\kf$ in the first Brillouin zone, one needs to focus on the little group $\mathfrak{G}_{\kf}$ of the respective k-point, which includes all symmetries $\{ R|\mathbf{v} \}$ satisfying $\kf = \kf^\prime = R \kf+\boldsymbol{G}$, where $\mathbf{G}$ is a vector of the reciprocal lattice. If $H_0$ is invariant under a spatial symmetry $h$, then from the above relation it follows that at all k-points with $h \in \mathfrak{G}_{\kf}$ the Bloch Hamiltonian matrix $\mathcal{H}_0(\kf)$ commutes with the band representation $\rho_{\mathfrak{G}}^{\kf}(h)$. From this commutative property it follows that eigenstates of $\mathcal{H}_{0}(k)$ transform as irreps of $\mathfrak{G}_{\kf}$, thus identifying these irreps is equivalent to determining symmetry properties of bands.

Now we analyze the Green's function of an interacting system. We assume a non-degenerate many-body ground state $\ket{0}$. It directly follows that the ground state is an eigenstate of every $U_h$. Since $U_h$ is unitary the eigenvalues must have modulus one and we can write for $h \in \mathfrak{G}$
\begin{equation}
U_h \ket{0} = e^{i \phi_h} \ket{0},
\end{equation}
with $\phi_h \in \left[ 0,2\pi \right) $. For imaginary time $\tau > 0$ we have
\begin{align}
G_{\alpha \beta}(\tau, \kf ) &=  -\bra{0} c_{\kf \alpha}(\tau) c^{\dagger}_{\kf \beta}\ket{0} \nonumber \\
&= -\bra{0} e^{H \tau}c_{\kf \alpha} e^{-H \tau} c^{\dagger}_{\kf \beta}  \ket{0} \nonumber \\
&= -\bra{0} e^{ U_h^{\dagger} H U_h \tau}c_{\kf \alpha} e^{-U_h^{\dagger} H U_h \tau} c^{\dagger}_{\kf \beta}  \ket{0}  \nonumber \\
&= -\bra{0} U_h^{\dagger} e^{H \tau} U_h c_{\kf \alpha} U_h^{\dagger} e^{-H \tau} U_h c^{\dagger}_{\kf \beta} U_h^{\dagger} U_h  \ket{0}  \nonumber\\
&= -\sum_{\gamma \delta} \rho_{\mathfrak{G}}^{\kf}(h)^*_{\alpha \gamma} \bra{0} c_{\kf'\gamma}(\tau ) c^{\dagger}_{\kf' \delta} \ket{0} \rho_{\mathfrak{G}}^{\kf}(h)_{\delta \beta} \nonumber \\
&= \sum_{\gamma \delta} \rho_{\mathfrak{G}}^{\kf}(h)^* _{\alpha \gamma} G_{\gamma \delta}(\tau, \kf' ) \rho_{\mathfrak{G}}^{\kf}(h)_{\delta \beta} . \label{gspat}
\end{align} 

A similar calculation holds for $\tau < 0$. Going to Matsubara frequencies i.e.~$G(i \omega, \kf )$ yields the same result for each value of $i \omega$. 
The above result shows, for a unique many-body ground state, that the Green's function transforms under the band representations in the same way as a (non-interacting) Bloch Hamiltonian.~\footnote{Note that for a degenerate ground state the physical system at $T= 0$ has a freedom in the choice of the ground state. This freedom can result in the spontaneous breaking of a symmetry of the Hamiltonian. In this case there is also an ambiguity in the Green's function in eq.~\eqref{defGt0}. This can lead to the Green's function not fulfilling the above symmetry relations. For non-spatial symmetries this possibility was already observed in Ref.~ \onlinecite{gurarie1d}.
However taking the zero-temperature limit from the finite temperature Matsubara Green's function in the case of a degenerate ground state, the resulting Green's function turns out to be an average over orthonormal basis states spanning the degenerate ground state space. Since the symmetry operators $U_h$ are unitary, the space of ground states is invariant under their action. Hence the Matsubara Green's function in the zero-temperature limit also commutes with the band representations of the spatial symmetries even for a degenerate ground state.} Similarly, provided the GNSC-conditions are fulfilled, a topological Hamiltonian $H_{\rm T}$ may be defined according to $H_{\rm T} = -G^{-1}(0,\kf)$, which transforms in the same way
\begin{equation}
\big( H_{\nk{T}}(\kf) \big)_{\alpha \beta} = \sum_{\gamma \delta} \rho_{\mathfrak{G}}^{\kf}(h)^*_{\alpha \gamma} \big( H_{\nk{T}}(\kf ') \big)_{\gamma \delta } \rho_{\mathfrak{G}}^{\kf}(h)_{\delta \beta} .
\label{hTbr}
\end{equation}

In particular, if $G( i \omega, \kf )$ fulfills the GNSC-conditions, $H_{\nk{T}}(\kf )$ is also continuous (even continuously differentiable) in $\kf$. It follows that the eigenvalues of $H_{\nk{T}}(\kf )$ form continuous bands in k-space that can be labeled by irreps of the little group of the respective k-points. Analogous to the non-interacting case, compatibility relations yield restrictions on the connectivity of the bands. This allows to write symmetry representations of bands as linear combinations of EBRs. Also in the case of TR present in the interacting system the resulting irreps of the little groups and EBRs have been classified and can be applied.~\cite{Bradlyn2017,EBRs}
We later clarify the meaning of assigning EBRs to the bands of a topological Hamiltonian in an interacting system.

\section{EBR-based analysis of the topological Hamiltonian: use and limitations}
\label{hteigv}

In this section, we discuss the interpretation of EBR-based analysis of the topological Hamiltonian in the spirit of TQC. 
For gapped non-interacting Hamiltonians, topological indices are invariant under unitary transformations of occupied (unoccupied) single-particle states. As a result, in terms of an EBR analysis we are concerned usually with the combined transformation properties of occupied single-particle states, independent of their energy ordering.~\cite{Bradlyn2017, EBRs}
To apply non-interacting classifications to $H_{\nk{T}} (\kf )$, an equivalent distinction is required. Following Refs.~\onlinecite{wang12inv,wang12}, for a Green's function fulfilling the GNSC-conditions, the eigenvalues of $H_{\nk{T}} (\kf )$ can be classified as either an L-zero or an R-zero.~\footnote{Originally the classification of R-zeros and L-zeros was used to discuss the eigenvalues of $G(0,\kf )$. This is equivalent to $H_{\nk{T}} (\kf )$, because both share the same eigenvectors.} An eigenvalue of the topological Hamiltonian $\mu_{\alpha }( \kf )$ is called an R-zero if $\mu_{\alpha } ( \kf )< 0 $ and an L-zero if $\mu_{\alpha }( \kf ) > 0 $. 
The R-zeros and L-zeros can each be written as a linear combination of EBRs with positive integer coefficients if they each correspond to a trivial set of bands or as a quasiband representation if they each correspond to a topological set of bands.
In a non-interacting system the R-zeros correspond to the occupied single-particle states and the eigenvalues $\mu_{\alpha }( \kf )$ are equal to the single-particle energies. For interacting systems, the eigenvalues can give some indication how the spectral weight is distributed on the real frequency axis.

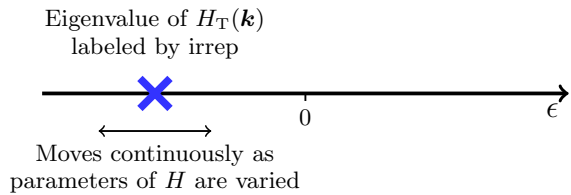
\begin{figure}
\begin{center}
\begin{tikzpicture}
\usetikzlibrary{shapes.misc}

\tikzset{cross/.style={cross out, draw=black, minimum size=2*(#1-\pgflinewidth), inner sep=0pt, outer sep=0pt, line width=1mm },
cross/.default={3mm}}

\tikzset{
      mynode/.style={align=center,text width=4cm},
      myarrow/.style={->, >=latex', shorten >=1pt, thick},
      mylabel/.style={text width=9em, text centered} 
    } 

\draw[->] [line width=0.5mm] (-3.5,0) -- (3.5,0) node[below left] {\large $\epsilon$};
\draw[-] [line width=0.25mm] (0,0) -- (0,-0.1) node[below] {$0$};
\node[mynode] at (-2,0.8) {Eigenvalue of $H_{\nk{T}} (\kf )$ labeled by irrep};
\draw[<->] [line width=0.25mm] (-2.75,-0.5) -- (-1.25,-0.5);
\node[mynode] at (-2.0,-1.0) {Moves continuously as \hbox{parameters} of $H$ are varied};

\draw (-2,0) node[cross,blue!80] {};

\end{tikzpicture}
\end{center}
\caption{Eigenvalues of the topological Hamiltonian at high symmetry k-points can be labeled by irreps of the little group of the k-point. The gap in the spectral function at $\omega = 0$ causes the eigenvalues to be real. Depending on the sign of the eigenvalue it is either an R-zero or an L-zero. As the parameters in the many-body Hamiltonian are varied, the eigenvalues move on the real axis as long as the gap is maintained. The gap further prevents the eigenvalues from crossing zero. By assumption we excluded a zero eigenvalue of $G(0,\kf)$, so that all eigenvalues of the topological Hamiltonian are finite. If also the Green's function does not break a symmetry while the parameters of the many body Hamiltonian are varied, the multiplicity of irreps of R-zeros cannot change.}
\label{fig:irrepsketch}
\end{figure}

Let us consider continuously changing the parameters of the many-body Hamiltonian, while maintaining symmetries and the ground state staying non-degenerate. We assume that $G(i \omega, \kf )$ and hence $G(0, \kf )$ also changes continuously on the path.  
For each point on this continuous path we can label the eigenvalues of the topological Hamiltonian at high symmetry k-points by the respective little group irreps, see also Fig.~\ref{fig:irrepsketch}. If the gap in the spectral function at zero frequency remains non-zero, then $G(0, \kf )$ remains Hermitian and the eigenvalues of the topological Hamiltonian move continuously on the real axis. Additionally the eigenvalues cannot cross over zero, because their absolute value is bounded from below if the gap in the spectral function remains non-zero as shown in Appendix \ref{anaG}. Further imposing the condition that $G(0, \kf ) $ never becomes singular on the path makes the inverse of $G(0, \kf ) $ well defined and finite. Hence the eigenvalues of the topological Hamiltonian also stay finite. Taking the above considerations together, at a high symmetry k-point the multiplicity of irreps of R-zeros (L-zeros) must be maintained under the above assumptions. The classification with respect to the EBR approach is therefore robust under continuous changes of the Green's function. 

The above considerations provide an interpretation to the Green's function invariants.  
For any ground state of an interacting Hamiltonian that can be adiabatically connected to a non-interacting Hamiltonian without the corresponding Green's function violating the GNSC-conditions, the Green's function invariants must match the non-interacting invariants obtained in the limit. 
Similarly, along any path on which the ground state is unique and whose endpoints are non-interacting models with ground states of different non-interacting topological indices, there must be at least one point where an L-zero becomes an R-zero and/or vice-versa. This requires at least one of the GNSC-conditions to be violated, which occur for three different scenarios:

(i) A gap closing in the spectral function at $\omega = 0$, which corresponds to a zero-energy excitation with finite quasiparticle weight. This is analogous to a change of a topological invariant by a gap closing in the non-interacting limit. For fixed filling, the L-zero and R-zero of different irreps must exchange at $\omega = 0$ in this scenario.

(ii) A zero eigenvalue in $G(0,\kf )$, corresponds to a divergence in the self-energy $\Sigma(i \omega ,\kf)$ at zero frequency, defined according to:
\begin{equation}
G(i \omega , \kf ) = \big( i \omega - \mathcal{H}_0 ( \kf ) - \Sigma(i \omega ,\kf) \big)^{-1}.
\end{equation}
This is only possible with interactions, as a non-interacting $G(0,\kf)$ cannot have zeros provided the energy spectrum is bounded. For the invariants defined previously in Refs.~\onlinecite{Volovik2003,Volovik2009He, Volovik2010,gurarieG,wang10,wang12inv,wang12,Wang_2013} the possibility of a change by a zero in the Green's function was recognized in Ref.~\onlinecite{gurarieG} and is discussed in Refs.~\onlinecite{gurarie1d,prlsshpu,budich_rev,unkonv_pert}. In the present context, this corresponds to an L-zero and R-zero of different irreps exchanging at infinity. 

(iii) A change in the symmetry of the many-body Hamiltonian or ground state. The latter case may occur via spontaneous symmetry breaking, which lowers the symmetry of both the ground state and Green's function. As a result, any Green's function invariants associated with the broken symmetries become ill-defined. See also the discussion Ref.~\onlinecite{gurarie1d} for the case of a chiral symmetry.
However, since the many-body Hamiltonian is invariant under the spontaneously broken symmetry this implies that the ground state must be degenerate. We have excluded this possibility by assumption in our analysis. 

At this point, we should note three caveats related to the above discussion and a possible correspondence of the 
Green's function invariants to SPT phases in the presence of interactions.

The first caveat, also discussed in Ref.~\onlinecite{Kivelson_2010}, is that there exist uniquely interacting SPT phases, which cannot be adiabatically connected to non-interacting limits provided certain symmetries are preserved. Within these phases, Green's function invariants obtained from the topological Hamiltonian are not constrained by the requirement of non-interacting correspondence.
In principle, they may take either any value when the GNSC-conditions are fulfilled, or they may not be well-defined.
An example of the latter case was recently demonstrated in Ref.~\onlinecite{iraola2021topological} for a 1D model exhibiting two gapped phases adiabatically connected to non-interacting limits, in addition to an interacting SPT phase. The latter was characterized by a divergence in the self-energy at zero frequency, i.e. a zero eigenvalue in $G(0,\kf)$ over the entire phase.

The second caveat is that the Green's function only probes single-particle excitations. As a result, the spectrum of $H_{\nk{T}} (\kf )$ may remain gapped with finite eigenvalues even as the spectrum of the many-body Hamiltonian becomes gapless with respect to a multi-particle excitation. This allows, in principle, for a transition between distinct SPT phases where the single-particle Green's function invariants do not change. 
On the other hand it is also possible that a transition with a gap closing of the spectral function gets replaced by a zero eigenvalue of $G(0,\kf )$ as discussed in Refs.~\onlinecite{gurarie1d,unkonv_pert}.

The third caveat is that the Green's function invariants may also change without a phase transition, i.e.~while the ground state remains non-degenerate. This applies to cases where the topological classification breaks down upon including interactions, as explicitly demonstrated in Ref.~\onlinecite{kitaev1D} for the CAZ symmetry class BDI where the $\mathbb{Z}$ classification breaks down to a $\mathbb{Z}_4$ ($\mathbb{Z}_8$ without particle number conservation). For the model in Ref.~\onlinecite{kitaev1D} it was argued that the Green's function invariant changes by a zero eigenvalue in the Green's function.~\cite{gurarie1d}  We expect the same to apply regarding an application of EBRs to the Green's function, especially for cases where EBRs and invariants may be explicitly related.

In conclusion we find that a precise statement in which cases there is a correspondence between SPT phases and an EBR analysis of the Green's function remains an important topic for future research.

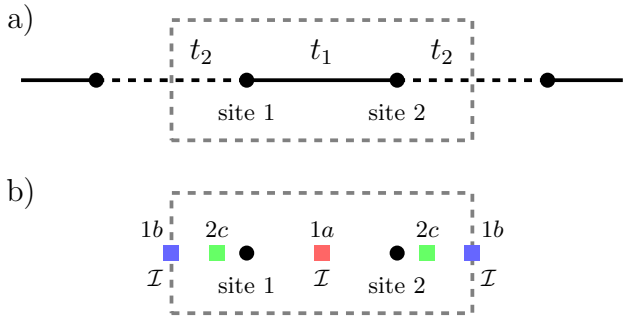
\begin{figure}
\begin{center}
\begin{tikzpicture}
\node at (0,1.8) {\large a)};
\draw [dashed,line width=0.5mm,black!50] (2,0.2) -- (6,0.2) -- (6,1.8) -- (2,1.8) -- (2,0.2);

\draw [line width = 0.5mm] (0,1) -- (1,1);
\draw [dashed,line width = 0.5mm] (1,1) -- (3,1);
\draw [line width = 0.5mm] (3,1) -- (5,1);
\draw [dashed,line width = 0.5mm] (5,1) -- (7,1);
\draw [line width = 0.5mm] (7,1) -- (8,1);

\node [above] at (2.4,1.1) {\large $t_2$};
\node [above] at (4,1.1) {\large $t_1$};
\node [above] at (5.6,1.1) {\large $t_2$};

\fill (3,1)  circle[radius=0.1]; \node [below] at (3,0.8) {site $1$};
\fill (5,1)  circle[radius=0.1]; \node [below] at (5,0.8) {site $2$};
\fill (1,1)  circle[radius=0.1]; 
\fill (7,1)  circle[radius=0.1]; 

\node at (0,-0.5) {\large b)};
\draw [dashed,line width=0.5mm,black!50] (2,-2.1) -- (6,-2.1) -- (6,-0.5) -- (2,-0.5) -- (2,-2.1);

\fill (3,-1.3)  circle[radius=0.1]; \node [below] at (3,-1.5) {site $1$};
\fill (5,-1.3)  circle[radius=0.1]; \node [below] at (5,-1.5) {site $2$};

\fill[red!60] (4-0.1, -1.3-0.1) rectangle (4+0.1,-1.3+0.1); \node [above] at (4,-1.2) {$1a$};
\fill[blue!60] (2-0.1, -1.3-0.1) rectangle (2+0.1,-1.3+0.1); \node [above left] at (2,-1.2) {$1b$};
\fill[blue!60] (6-0.1, -1.3-0.1) rectangle (6+0.1,-1.3+0.1); \node [above right] at (6,-1.2) {$1b$};
\fill[green!60] (2.6-0.1, -1.3-0.1) rectangle (2.6+0.1,-1.3+0.1); \node [above] at (2.6,-1.2) {$2c$};
\fill[green!60] (5.4-0.1, -1.3-0.1) rectangle (5.4+0.1,-1.3+0.1); \node [above] at (5.4,-1.2) {$2c$};
\node [below left] at (2, -1.3-0.1) {$\mathcal{I}$};
\node [below right] at (6, -1.3-0.1) {$\mathcal{I}$};
\node [below] at (4, -1.3-0.1) {$\mathcal{I}$};

\end{tikzpicture}

\end{center}
\caption{a) The SSH model with the unit cell, sites and hoppings shown, b) Unit cell of the SSH model with Wyckoff positions 1a,1b and 2c. The centers of the inversion symmetry $\mathcal{I}$ are located in 1a and 1b. See the text for the description of the model.}
\label{fig::unitcell}
\end{figure}

\section{An example: EBRs for the Green's function in the SSH+U model}
\label{demo}

In this section we give a detailed demonstration of the above extension of EBRs for the Green's function on the simple SSH+U model in the context of inversion symmetry and discuss the correspondingly diagnosed topological phases.

\subsection{The SSH+U model}

The SSH+U model is defined by:
\begin{align}
H =& t_1  \sum_{j \sigma} (c^{\dagger}_{j2 \sigma}c_{j1 \sigma} + c^{\dagger}_{j1\sigma}c_{j2\sigma}) \nonumber \\
&+ t_2 \sum_{j \sigma} (c^{\dagger}_{j+1,1\sigma}c_{j2\sigma} + c^{\dagger}_{j2 \sigma}c_{j+1,1 \sigma}) \nonumber \\
&+ U \sum_{j \alpha} (n_{j \alpha \uparrow} - \frac{1}{2}) (n_{j \alpha \downarrow} - \frac{1}{2}) .
\label{H_sshpu}
\end{align}
where $c^{\dagger}_{j \alpha \sigma}$ ($c_{j \alpha \sigma}$) creates (annihilates) an electron in unit cell $j$, site $\alpha \in \{1,2\}$ and with spin $\sigma$. We consider the model at half-filling.
The form of the interaction is such that the chemical potential is zero and thus included in the Hamiltonian. The unit cell with the hopping parameters is shown in Fig.~\ref{fig::unitcell}. We also show the Wyckoff positions and the centers of inversion symmetry. 

The model belongs to the CAZ symmetry class BDI, which implies\cite{gurarieG} there is an integer topological classification for the Green's function, with a topological invariant $N_1$.
A DMRG study on a finite system~\cite{gurarie1d} investigated the case $t_1 + t_2 > 0$ and found the system to be trivial with $N_1 = 0$ for $t_1 -t_2 > 0$ while for $t_1 -t_2 < 0$ the system is in a topological phase with $N_1 = 2$ for all values of $U> 0$.~\cite{gurarie1d} At the transition at $t_1 = t_2$ the model reduces to the 1D Hubbard model. While for $U=0$ the system is in a metallic phase with band crossing at $k= \pi $. For any $U>0$ the system is a Mott insulator with a charge gap and gapless spin excitations.~\cite{lieb_1d_hubb} Since collective excitations are not visible in the single-particle Green's function the transition in the bulk Green's function topological invariant $N_1$ happens by a zero in the Green's function at $k=\pi$.~\cite{prlsshpu} In the following, we investigate how this transition at finite $U$ is related to spatial inversion symmetry. 

\subsection{Symmetry analysis}

The model possesses time-reversal (TR), particle-hole (PH) and chiral symmetries (CS). The latter can be associated with the following representation for the Green's function in k-space:~\cite{gurarie1d}
\begin{equation}
U_{\nk{CS}} = \begin{pmatrix} 
1 & 0  \\
0 &  -1 \\ 
\end{pmatrix} = \sigma_3 .
\end{equation}
Here, it is not necessary to discuss the spin; since the model is also invariant under spin rotations. We suppressed the spin indices throughout the following. 
The CS places a restriction on the topological Hamiltonian:
\begin{equation}
H_{\nk{T}}(k) = - \sigma_3 H_{\nk{T}}(k)  \sigma_3 .
\label{htc_s3}
\end{equation}
We expand the topological Hamiltonian into Pauli matrices. Because of eq.~\eqref{htc_s3} only terms proportional to $\sigma_1$ and $\sigma_2$ are allowed. So we can write
\begin{equation}
H_{\nk{T}}(k) = q_1(k) \sigma_1 + q_2(k)\sigma_2.
\label{pauliexp}
\end{equation}
$q_1(k)$ and $q_2(k)$ are real, k-dependent coefficients. Setting $q(k) = q_1(k) + i q_2(k)$, the Green's function topological invariant $N_1$ can then be written in terms of the topological Hamiltonian as~\cite{gurarieG,gurarie1d}
\begin{align}
N_1 &= 2 ~ \nk{tr} \int \frac{ \nk{d}k}{4 \pi i } U_{\nk{CS}} H_{\nk{T}}(k) \partial_k H_{\nk{T}}^{-1}(k) \nonumber \\
&= 2 \int \frac{ \nk{d}k }{2 \pi i } q(k) \partial_k q^{-1}(k). \label{defN1}
\end{align}
The factor of two in front of the integral comes from the spin degeneracy in the model. The trace goes over the matrix indices of the topological Hamiltonian. The invariant just measures how often $q(k)$ winds around the origin in the complex plane.

We consider the inversion centered in the Wyckoff position 1a of the unit cell. The high-symmetry $k$-points are $\kappa = 0$ and $\kappa = \pi$, whose little groups contain inversion and the identity. For these $k$-points, the electron operators transform as:
\begin{align}
\mathcal{I} c^{\dagger}_{\kappa \alpha \sigma} \mathcal{I}^{\dagger} &= \sum_{\beta} \big( \rho_{\mathfrak{G}}^{\kappa}(\mathcal{I}) \big)_{\beta \alpha} c^{\dagger}_{\kappa \beta \sigma}
\end{align}
with the band representation of the inversion operator $\rho_{\mathfrak{G}}^{\kf}(\mathcal{I})$ given by
\begin{equation}
\rho_{\mathfrak{G}}^{\kf}(\mathcal{I}) = \begin{pmatrix} 
0 & 1  \\
1 &  0 \\ 
\end{pmatrix} = \sigma_1 .
\label{inv_rep}
\end{equation}
At these high-symmetry k-points, the eigenvectors of the topological Hamiltonian may be labeled by the eigenvalue of the inversion symmetry, which can be $+1$ or $-1$. In this case working with the inversion eigenvalues is equivalent to work with irreps, because there are only two irreps which can be distinguished by the eigenvalue of inversion.
With only inversion symmetry there exist four possible EBRs each containing one band (two degenerate bands upon including TR symmetry/degeneracy in spin space).
This provides four equivalence classes for the Green's function, defined by the EBR of the lower band, as summarized in Table~\ref{res_top}. In the 1D group $P\bar{1}$, the EBRs are induced by either an s-like (even under inversion) or a p-like Wannier function (odd under inversion) in the 1a or the 1b Wyckoff position. We label the EBR by the orbital type in the respective Wyckoff position which induces the EBR. See also Table~\ref{res_top} for all possible EBRs and band structures. In the SSH+U model where both chiral and inversion symmetry are present the index $N_1$ can be directly related to the inversion eigenvalues at $k =0,\pi$: $N_1 = 4n +2$ with $n \in \mathbb{Z}$ if both inversion eigenvalues have opposite sign and $N_1 = 4n $ if the signs are equal (see Appendix \ref{app:ssh_help}). 

\begin{table}
\caption{\label{res_top}Four possibilities to label the bands of the topological Hamiltonian by EBRs in the SSH+U model. The inversion eigenvalues at $k = 0$ and $k= \pi$ fully determine the irreps. The EBR is labeled by the orbital type which induces the EBR. The inversion eigenvalues are also indicated in the band structure sketches.}
\begin{ruledtabular}
\begin{tabular}{ccccc}
\specialcell[c]{Lower irrep \\ at $k=0$} & \specialcell[c]{Lower irrep \\ at $k=\pi$} & \specialcell[c]{Lower \\ EBR} &  $N_1$ & \specialcell[c]{Band structure \\ sketch}  \\
\hline
$\Gamma_+$      & $X_+$    & s\ti{1a}             & $0$   &  \begin{tikzpicture} \def\x{0.4} \draw (0,\x) --(1,\x); \draw [fill] (0,\x) circle [radius=0.05]; \node [left] at (0,\x) {$-$}; \draw [fill] (1,\x) circle [radius=0.05]; \node [right] at (1,\x) {$-$}; \draw (0,0) --(1,0); \draw [fill] (0,0) circle [radius=0.05]; \node [left] at (0,0) {$+$}; \draw [fill] (1,0) circle [radius=0.05]; \node [right] at (1,0) {$+$}; \node [below] at (0,0) {$0$}; \node [below] at (1,0) {$\pi$}; \end{tikzpicture}   \\
\hline
$\Gamma_-$ & $X_-$    & p\ti{1a}              & $0$    &  \begin{tikzpicture} \def\x{0.4} \draw (0,\x) --(1,\x); \draw [fill] (0,\x) circle [radius=0.05]; \node [left] at (0,\x) {$+$}; \draw [fill] (1,\x) circle [radius=0.05]; \node [right] at (1,\x) {$+$}; \draw (0,0) --(1,0); \draw [fill] (0,0) circle [radius=0.05]; \node [left] at (0,0) {$-$}; \draw [fill] (1,0) circle [radius=0.05]; \node [right] at (1,0) {$-$}; \node [below] at (0,0) {$0$}; \node [below] at (1,0) {$\pi$}; \end{tikzpicture}       \\
\hline
$\Gamma_+$       & $X_-$    & s\ti{1b}            & $2$    &  \begin{tikzpicture} \def\x{0.4} \draw (0,\x) --(1,\x); \draw [fill] (0,\x) circle [radius=0.05]; \node [left] at (0,\x) {$-$}; \draw [fill] (1,\x) circle [radius=0.05]; \node [right] at (1,\x) {$+$}; \draw (0,0) --(1,0); \draw [fill] (0,0) circle [radius=0.05]; \node [left] at (0,0) {$+$}; \draw [fill] (1,0) circle [radius=0.05]; \node [right] at (1,0) {$-$}; \node [below] at (0,0) {$0$}; \node [below] at (1,0) {$\pi$}; \end{tikzpicture}    \\
\hline
$\Gamma_-$       & $X_+$    & p\ti{1b}            & $2$    &  \begin{tikzpicture} \def\x{0.4} \draw (0,\x) --(1,\x); \draw [fill] (0,\x) circle [radius=0.05]; \node [left] at (0,\x) {$+$}; \draw [fill] (1,\x) circle [radius=0.05]; \node [right] at (1,\x) {$-$}; \draw (0,0) --(1,0); \draw [fill] (0,0) circle [radius=0.05]; \node [left] at (0,0) {$-$}; \draw [fill] (1,0) circle [radius=0.05]; \node [right] at (1,0) {$+$}; \node [below] at (0,0) {$0$}; \node [below] at (1,0) {$\pi$}; \end{tikzpicture}      \\
\end{tabular}
\end{ruledtabular}
\end{table}

\begin{figure}
\begin{center}
\begin{tikzpicture}[scale=0.6]

\definecolor{r1}{RGB}{255, 170 , 170}
\definecolor{b2}{RGB}{120, 160, 255}
\definecolor{g3}{RGB}{120, 255, 160}
\definecolor{g4}{RGB}{255,255,150}

\fill[r1] (-5,-5)--(0,0)--(-5,5);
\fill[b2] (-5,-5)--(0,0)--(5,-5);
\fill[g3] (5,5)--(0,0)--(-5,5);
\fill[g4] (5,5)--(0,0)--(5,-5);

\draw[->] [line width=0.45mm] (-5,0) -- (5,0) node[below left] {\large $t_1$};
\draw[->] [line width=0.45mm] (0,-5) -- (0,5) node[below left] {\large $t_2$};

\draw [line width=0.25mm] (-5,-5) -- (5,5);\node[below right] at (2.5,2.5) { $t_1 = t_2$};
\draw [line width=0.25mm] (5,-5) -- (-5,5); \node[below left] at (-2.5,2.5) { $t_1 = -t_2$};

\node[above] at (-3,-1) {\large \textbf{s\ti{1a}}};
\node[below] at (-3,-1) {$N_1 = 0$};

\node[above] at (3,-1) {\large \textbf{p\ti{1a}}};
\node[below] at (3,-1) {$N_1 = 0$};

\node[above] at (-1,3) {\large \textbf{p\ti{1b}}};
\node[below] at (-1,3) {$N_1 = 2$};

\node[above] at (-1,-3) {\large \textbf{s\ti{1b}}};
\node[below] at (-1,-3) {$N_1 = 2$};

\end{tikzpicture}
\end{center}
\caption{Phase diagram obtained from an EBR analysis for the Green's function of the SSH+U model. The phases are labeled by the EBR lowest in energy, derived from the inversion eigenvalues (see Table~\ref{res_top}). The respective value of $N_1$ is also shown. The Green's function EBRs are independent of the value of the Hubbard interaction $U$, hence yielding the same phase diagram for any value of $U$.}
\label{ssh_pd}
\end{figure}
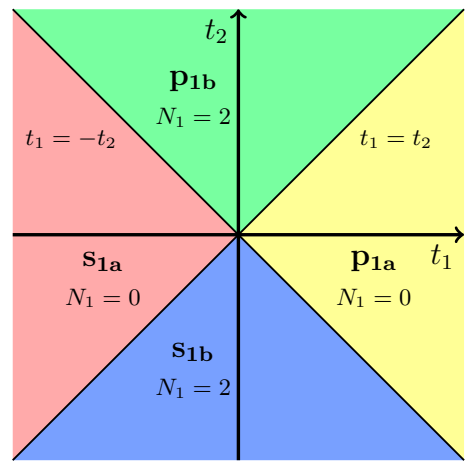

\begin{figure*}
\centering
 \begin{tabular}{cc}
  \includegraphics[width=0.49\linewidth]{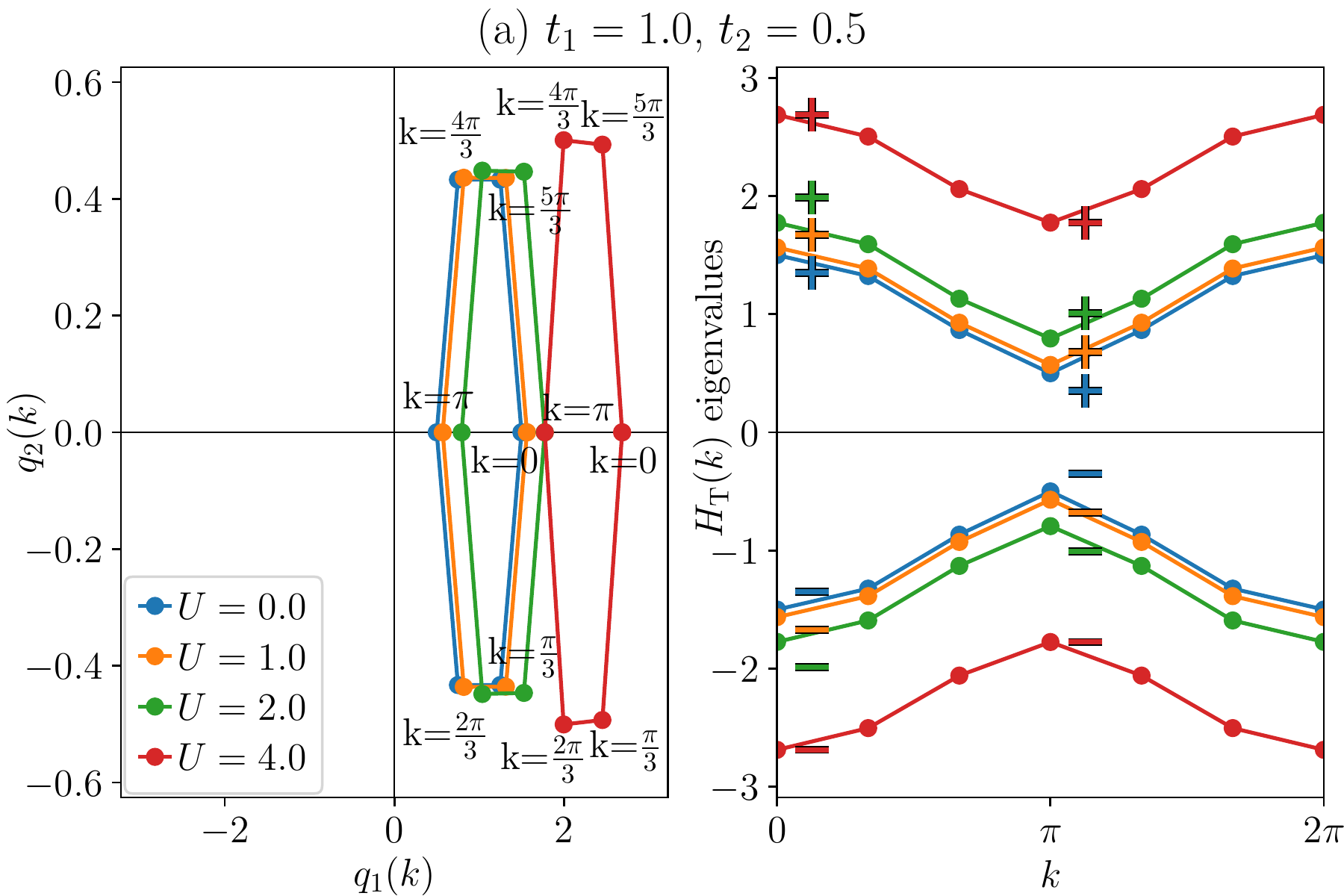} & \includegraphics[width=0.49\linewidth]{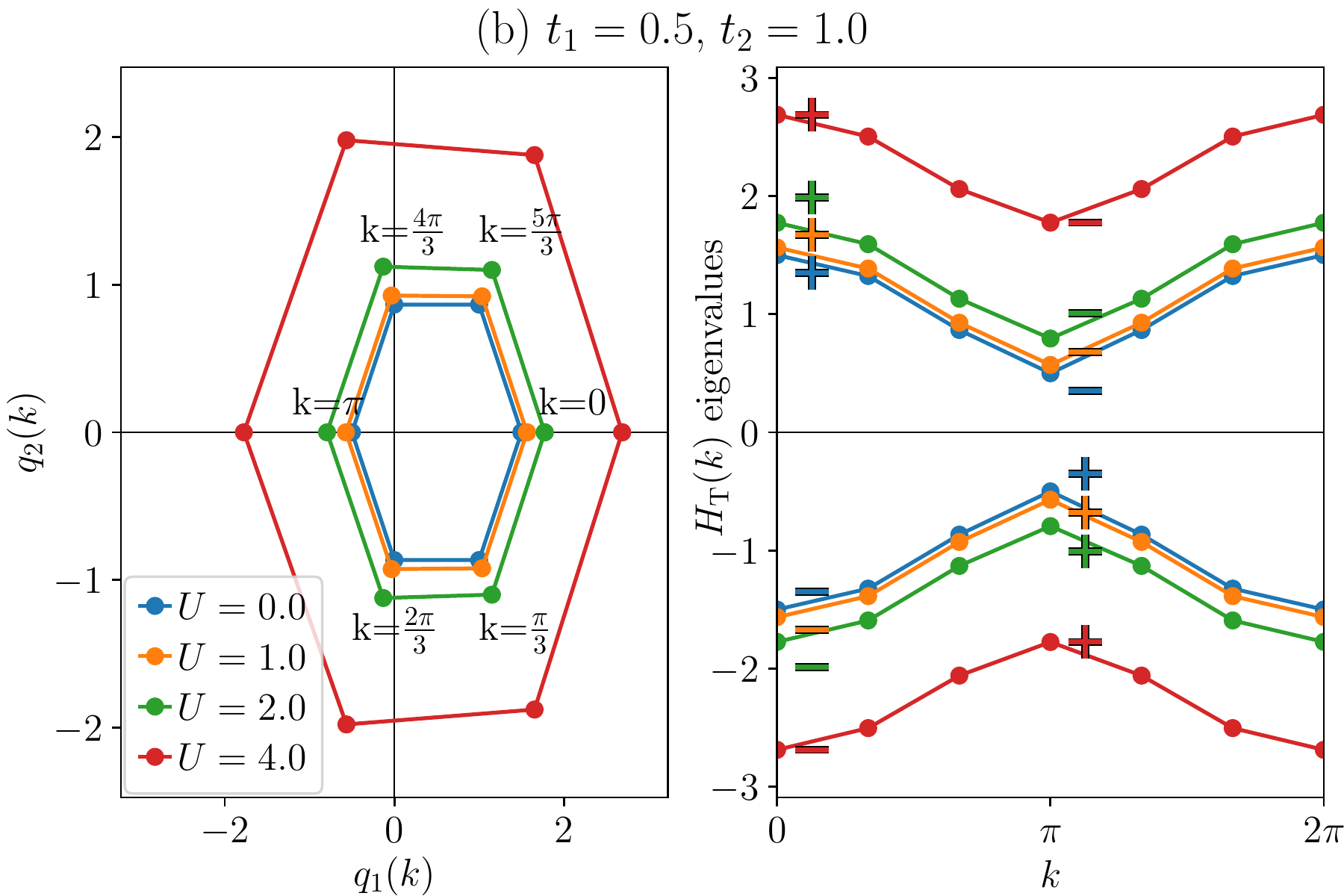} 
\end{tabular}
\caption{\label{uvary_4ph} Exact diagonalization results for the topological Hamiltonian as a function of $U$. Each figure corresponds to a representative choice of the hopping parameters $t_1$ and $t_2$ for each topological phase. Labeled by the EBR of the lower band the phases shown are (a) p\ti{1a}, (b) p\ti{1b}. The respective left plot shows the winding around the origin of the Pauli matrix expansion coefficients $q_1(k)$ and $q_2(k)$ (see eq.~\eqref{pauliexp}) when $k$ sweeps the Brillouin zone. The respective right plot shows the eigenvalues of the topological Hamiltonian together with the inversion eigenvalues of the respective eigenstates of the topological Hamiltonian at the high-symmetry k-points denoted by $+$ or $-$. The inversion eigenvalues fully determine the irreps. Increasing the Hubbard interaction $U$ enlarges the expansion coefficients $q_1(k)$ and $q_2(k)$ without changing the topology. }
\end{figure*}

\subsection{Exact diagonalization results}

To compute $H_T$ at finite $U$, we employ exact diagonalization (ED) calculations with six unit cells and periodic boundary conditions (PBC). In this context, a significant advantage of employing the irreps at high-symmetry $k$-points to characterize the topology of the Green's function is that they are well-defined in finite-size calculations without extrapolation. Nonetheless, a previous DMRG study that considered systems consisting up to 125 unit cells also did not find any indication for a finite size effect for the invariant $N_1$.~\cite{gurarie1d}

In ED, the Green's function is obtained by evaluating the Lehmann representation eq.~\eqref{lehmann}. Therefore we calculate the exact ground state in the $N$ particle sector ($N$ corresponding to half-filling) and the $m_{\nk{max}}$ lowest in energy exact eigenstates in the $N+1$ and $N-1$ particle sector of the full many-body Hamiltonian with Lanczos method. The number $m_{\nk{max}}$ is determined such that we at least take into account $99 \% $ of the spectral function in the respective Green's function entry. Since we are interested only in $G(0,k) $ and the Lehmann representation being essentially a (finite) pole expansion, this small neglect of spectral function has no influence on the EBR classification in the present system. 
From $G(0,k) $ the topological Hamiltonian is simply obtained by matrix inversion (see eq.~\eqref{htopdef}). The topological Hamiltonian is then analyzed in terms of EBRs and the invariant $N_1$. The resulting phase diagram from this analysis is shown in Fig.~\ref{ssh_pd}.

\begin{figure*}
\centering
 \begin{tabular}{cc}
  \includegraphics[width=0.49\linewidth]{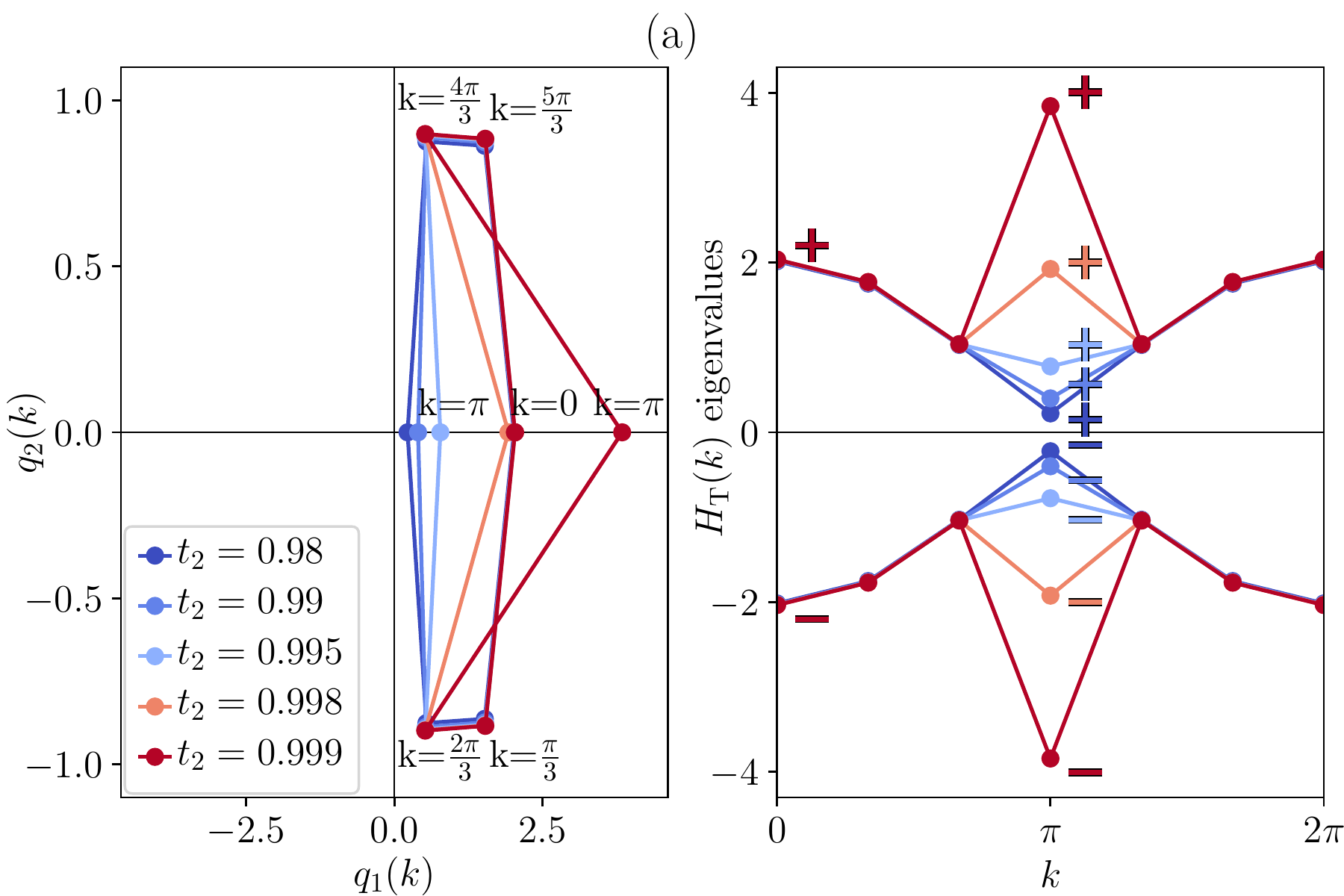} & \includegraphics[width=0.49\linewidth]{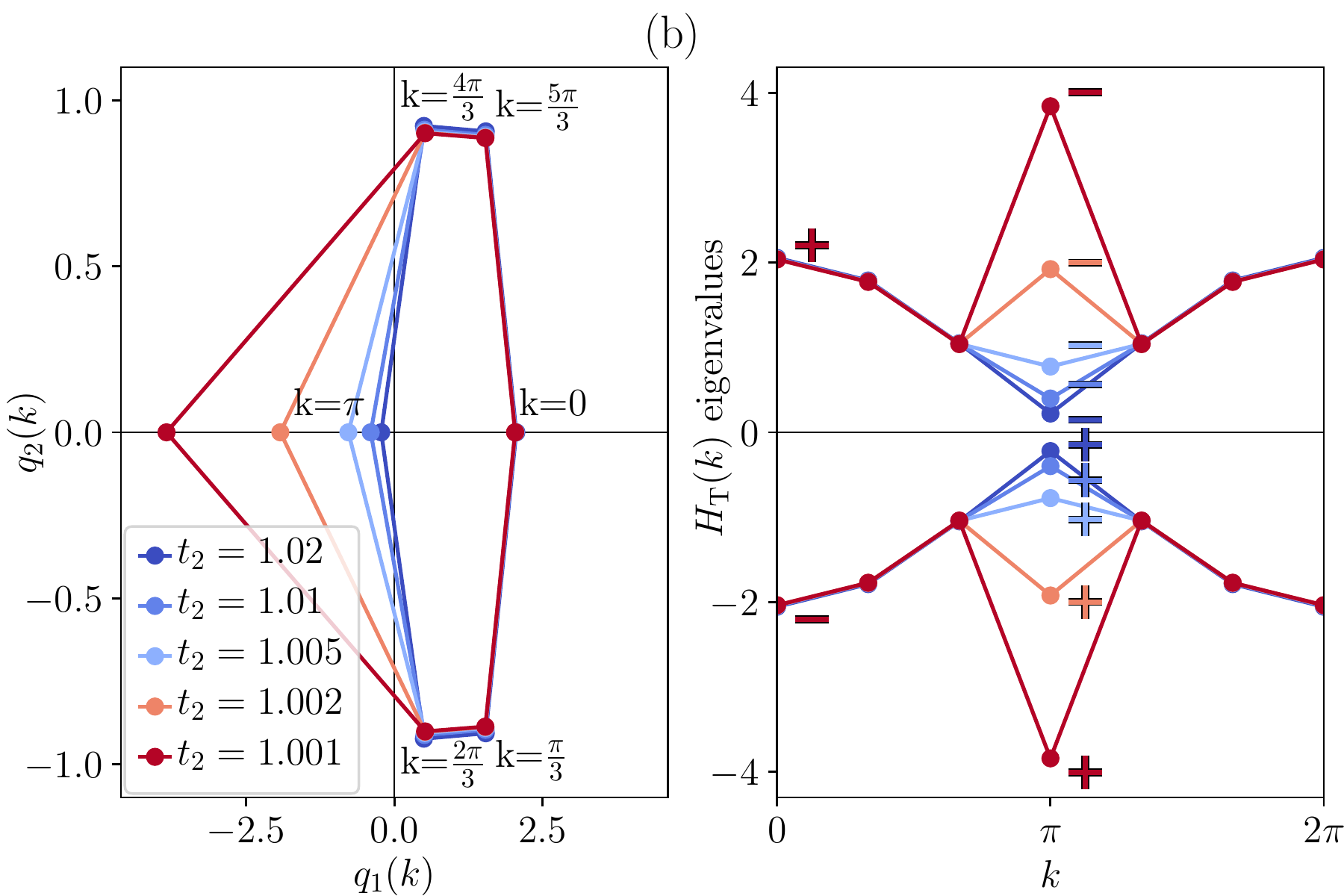} 
\end{tabular}
\caption{\label{ptrans} Exact diagonalization results for the topological Hamiltonian near the line $t_1 = t_2$. In each case, $t_1 =1$ and $U=1$ are fixed. The respective left figure shows the winding around the origin of the Pauli matrix expansion coefficients $q_1(k)$ and $q_2(k)$ (see eq.~\eqref{pauliexp}) when $k$ sweeps the Brillouin zone. The respective right plot shows the eigenvalues of the topological Hamiltonian together with the inversion eigenvalues of the respective eigenstates of the topological Hamiltonian at the high-symmetry k-points denoted by $+$ or $-$. The inversion eigenvalues fully determine the irreps. The results indicate that the transition happens by a swap over infinity of the eigenvalues of the topological Hamiltonian labeled by the inversion eigenvalues.}
\end{figure*}

Let us focus on the cases where $t_1, t_2 > 0$, which yields two phases: $t_1 > t_2$ corresponds to a lower band with EBR $p_{1a}$, while $t_1 < t_2$ corresponds to a lower band with EBR $p_{1b}$. The topological Hamiltonian and its eigenvalues together with the inversion eigenvalues are shown in Fig.~\ref{uvary_4ph}. For each phase we chose representative parameters $t_1$ and $t_2$, while increasing $U$. The results illustrate that a winding of the topological Hamiltonian in the $q_1$-$q_2$-plane around the origin i.e. $N_1 = 2$ corresponds to the inversion eigenvalues of the lower band having opposite sign at $k=0$ and $k= \pi$. The inversion eigenvalues of the lower band having the same sign corresponds to $N_1=0$.
Increasing $U$ enlarges the matrix elements of the topological Hamiltonian and hence also its eigenvalues. Intuitively this occurs because enlarging $U$ enlarges the gap in the spectral function. From the spectral representation, one expects that the matrix elements of $G(0, \kf ) $ become smaller. However in the present system $U$ has no influence on the inversion eigenvalues or $N_1$.

We now investigate what happens at the transition $t_1 = t_2$. For $U = 0$, this transition occurs via gap closure at $k=\pi$, at which point eigenvalues labeled $+$ and $-$ swap by crossing over zero. This behavior may be contrasted with finite $U$, which we investigate in Fig.~\ref{ptrans}.
Approaching $t_1 = t_2$, we find that the topological Hamiltonian and hence the self-energy at zero frequency starts to diverge at $k=\pi$, implying that the Green's function at zero frequency becomes zero. 
At the transition the eigenvalues of the topological Hamiltonian labeled by $+$ and $-$ swap by crossing over infinity. Thus the Green's function EBR classification changes by violating the condition that $G(0,\kf)$ must be non-singular i.e. condition two in \textit{Definition}~\ref{GNSCc}.
This agrees with the finding that the simultaneous transition of $N_1$ at $t_1 = t_2$ happens by a divergence in $\Sigma(0,\pi)$.~\cite{prlsshpu}
Note that one has to tune $t_2$ very close to the transition to observe the divergent behavior in the eigenvalues of the topological Hamiltonian. Initially it might look like the eigenvalues swap by crossing over zero, implying a gap closing in the spectral function at zero frequency (violating condition one in \textit{Definition}~\ref{GNSCc}) like in the non-interacting system.

\section{Conclusions} 
\label{Conclusions}

In this work we have investigated
the applicability of elementary band representations in the spirit of TQC and symmetry indicators to diagnose spatial- and time-reversal-symmetry protected topological phases in interacting insulators in terms of their single-particle Green's functions.

Starting from the fact that spatial symmetries enrich the topological classification of the Green's function, and provided there exists only a unique ground state of the interacting system, 
we illustrated that it is possible to define EBRs for the Green's function via the topological Hamiltonian in eq.~\eqref{htopdef}, in analogy to previously defined Green's function invariants~\cite{Volovik2003,Volovik2009He, Volovik2010,gurarieG,wang10,wang12inv,wang12,Wang_2013} imposed by the symmetries in the CAZ symmetry classes. 
We further established that the Green's function EBR classification can only change by (i) a gap closing in the spectral function at zero frequency, (ii) the Green's function becoming singular at zero frequency (i.e. $\det \big(G(0, \kf ) \big) = 0$) or (iii) the Green's function breaking a protecting symmetry. However, the question in which cases there is a strict correspondence
between an EBR classification of Green's functions and SPT phases remains a topic for future research.

As an example, we demonstrated the use of the EBRs for Green's functions on the SSH+U model, which is in
the CAZ symmetry class BDI and has spatial inversion symmetry. 
This model features a transition for $U>0$ where the Green's function becomes singular at zero frequency, which allows the eigenvalues of the topological Hamiltonian at high-symmetry k-points labeled by the inversion symmetry to swap by crossing over infinity.
Although we demonstrated the usage of EBRs for the Green's function only in one dimension, a similar analysis can also be applied in higher dimensions.

For numerical finite-size calculations on interacting models, the EBR evaluation may often prove valuable as it requires only the calculation of irreps of the topological Hamiltonian at a few high-symmetry $k$-points, and thus does not require integrations over $k$-space or explicit extrapolations to the thermodynamic limit.

\section*{Acknowledgements}
We thank Thomas Mertz, Titus Neupert, Frank Pollmann and J. L. Ma\~{n}es for helpful discussions. 
D.L. and R.V. acknowledge the Deutsche Forschungsgemeinschaft (DFG, German Research Foundation) for funding through Grant No. TRR 288 - 422213477 (Project B05).
M.G.V. and  M.I. acknowledge support from the Spanish  Ministerio de Ciencia e Innovacion (Grants number PID2019-109905GB-C21 and PGC2018-094626-B-C21) and Basque Government (Grant IT979-16).
Part of the work of M.G.V., S.M.W. and R.V. was carried out at Kavli Institute of Theoretical Physics (KITP), which is supported by the National Science Foundation under FQ 581 Grant No. NSF PHY-1748958.

\appendix

\section{Analytic properties of the Matsubara Green's function} \label{anaG}

In this Appendix we investigate analytic properties of the Green's function, necessary to apply EBRs to interacting systems in terms of the Green's function. For completeness and to clarify our notation we first review the definition and the spectral representation of the Matsubara Green's function. We then proof that, for a Green's function fulfilling the first two GNSC-conditions, the corresponding topological Hamiltonian is Hermitian with finite eigenvalues whose absolute value is bounded from below.

For simplicity we consider a lattice model defined on a basis of exponentially localized, orthonormal Wannier functions $ \phi_{i \alpha} (\rf ) $ consistent with the symmetries of our system, where the index $i$ labels the unit cell with lattice vector $\f{R}_i$ to which $ \phi_{i \alpha}(\rf ) $ belongs to. The Wannier function in that cell is specified by the index $\alpha$ which can also include spin. 
From the Wannier functions $\phi_{i\alpha}(\rf )$ one can construct Bloch-like wave functions $\psi_{\kf \alpha}(\rf )$ by a Fourier transform. 
In the basis of Bloch-like wave functions the single-particle Matsubara Green's function in the zero-temperature limit is defined as
\begin{equation}
G_{\alpha\beta}(\tau, \kf) = -\bra{0}\mathcal{T} c_{\kf \alpha}(\tau) c^{\dagger}_{\kf \beta}\ket{0},
\label{defGt0}
\end{equation}
where $c^{\dagger}_{\kf \alpha}$ ($c_{\kf \alpha}$) creates (annihilates) an electron in the Bloch-like state with crystal momentum $\kf$ and orbital index $\alpha$. The time evolution is in imaginary time $\tau$.
$\mathcal{T}$ denotes time ordering in imaginary time and $\ket{0}$ denotes the ground state, which in the following we assume to be non-degenerate. Here and throughout this paper we include the chemical potential $\mu$ in the Hamiltonian.~\footnote{Note that naively the chemical potential in an interacting insulator can be placed anywhere inside the gap of the spectral function. This may lead to ambiguities in the topological classification of $G$ near phase boundaries where the single-particle gap does not close. However even in the zero-temperature limit $\mu ( T \to 0 )$ can be uniquely defined by ensuring fixed particle density as a function of $T$. See also the discussion in Ref.~\onlinecite{mut0_kane}. We adopt this definition throughout.}
Going to frequency space we can write the Green's function in the Lehmann representation
\begin{align}
G_{\alpha \beta } (i \omega, \kf ) =& \sum_m  \Biggr[ \frac{ \bra{0}c_{ \kf \alpha} \ket{m} \bra{m} c^{\dagger}_{ \kf \beta} \ket{0}}{i \omega - (E_m - E_0)} \nonumber \\
&+ \frac{ \bra{m}c_{ \kf \alpha} \ket{0} \bra{0} c^{\dagger}_{ \kf \beta} \ket{m}}{i \omega + (E_m - E_0)} \Biggr] .
\label{lehmann}
\end{align}
For the ground state having $N$ particles the sum runs over the exact eigenstates $\ket{m}$ of the many-body Hamiltonian with $N+1$ and $N-1$ particles. The $E_m$ are the corresponding exact energy eigenvalues. $E_0$ is the ground state energy. 
Note that in the zero-temperature limit the discrete Matsubara frequencies $\omega_n$ become continuous $i \omega_n \to i \omega$. 
For the special case of a non-interacting Hamiltonian the Matsubara Green's function can be simply written in terms of the corresponding Bloch Hamiltonian, see eq.~\eqref{G0def}. 

The Matsubara Green's function $G(i \omega, \kf ) $ has an analytical extension from the imaginary axis to the whole complex plane except the real axis, i.e. $G(z,\kf)$ with $z \in \mathbb{C} \setminus \mathbb{R}$. From the Lehmann representation one obtains that $G(z,\kf)$ has poles on the real axis. 
We can write down the spectral representation for $G(z,\kf)$
\begin{equation}
G(z,\kf) = \int_{- \infty}^{\infty} \frac{\diff \omega' }{2 \pi} \frac{A(\omega', \kf)}{z - \omega'} 
\label{Gz}    
\end{equation}
with the matrix elements $A_{\alpha \beta}(\omega, \kf)$ of the spectral function given by 
\begin{align}
A&_{\alpha \beta}(\omega, \kf) \nonumber \\
=& \sum_n \bra{0}c_{\kf \alpha } \ket{n} \bra{n} c^{\dagger}_{\kf \beta} \ket{0} 2 \pi \delta \big(\omega- (E_n -E_0) \big) \nonumber \\
&+ \sum_n \bra{n}c_{\kf \alpha } \ket{0} \bra{0} c^{\dagger}_{\kf \beta} \ket{n} 2 \pi \delta \big( \omega- (E_0 -E_n) \big),
\label{spectralf}
\end{align}
where $\omega \in \mathbb{R}$. Inserting eq.~\eqref{spectralf} into eq.~\eqref{Gz} reproduces the Lehmann representation in eq.~\eqref{lehmann}. 
For the case of an infinite crystal the poles of $G(z,\kf)$ become dense and form a branch cut.~\cite{Luttana}
It can be shown from its definition that the spectral function is Hermitian and positive semi-definite for all $\kf$ and $\omega$. Further for a vector $\f{a}$ with the same dimension as the spectral function and $||\f{a}|| = 1$ the spectral function is normalized in the following sense
\begin{align}
& \int_{- \infty}^{ \infty} \frac{\diff \omega }{2 \pi}  \sum_{\alpha, \beta}  \overline{a}_{\alpha } A_{\alpha, \beta}(\omega, \kf) a_{\beta }  \nonumber \\
&= \sum_{\alpha \beta} \overline{a}_{\alpha } \bra{0} \lbrace c_{\kf \alpha}, c^{\dagger}_{\kf \beta} \rbrace \ket{0} a_{\beta } \nonumber \\
&= \sum_{\alpha, \beta} \overline{a}_{\alpha } a_{\beta } \delta_{\alpha \beta} \nonumber \\
&=1, 
\label{Anorm}
\end{align}
where $\overline{a}_{\alpha }$ denotes the complex conjugate of $a_{\alpha }$. 
Every complex matrix can be decomposed into a Hermitian and an anti-Hermitian part. Since $A(\omega, \kf )$ is Hermitian and with $z = x + i y$ with $x,y \in \mathbb{R}$ we can write the Green's function as
\begin{align}
G(z,\kf) =& \int_{- \infty}^{\infty} \frac{d \omega'}{2\pi} A(\omega', \kf)\frac{(x - \omega')}{(x - \omega')^2 + y^2} \nonumber \\
&- i \int_{- \infty}^{\infty} \frac{d \omega'}{2\pi} A(\omega', \kf)\frac{y}{(x - \omega')^2 + y^2}, \label{g1g2}
\end{align}
which defines a decomposition $G = G_1 + i G_2$ with both $G_1$ and $G_2$ being Hermitian.
Evaluating the limit of z approaching the real axis we must distinguish between taking the limit coming from the upper or the lower complex plane. For $\eta >0$ and $\omega \in \mathbb{R}$ one obtains
\begin{align}
\lim_{\eta \rightarrow 0} G(\omega \pm i \eta ,\kf) 
&=  \nk{PV} \int_{- \infty}^{\infty} \frac{d \omega'}{2\pi} \frac{A(\omega', \kf)}{(\omega  - \omega')} \mp \frac{i}{2}  A(\omega, \kf) ,
\label{gaherm}
\end{align}
where $\nk{PV}$ denotes the Cauchy principal value.

We now focus on the case where the spectral function has a non-zero gap at zero frequency for every k-point, i.e. the first condition in \textit{Definition}~\ref{GNSCc}. This causes both limits \hbox{$\lim_{\eta \rightarrow 0} G(\omega \pm i \eta ,\kf)$} to coincide within the gap and as a consequence $G(i \omega, \kf)$ is analytic in $i \omega$. Further in this case $G(0, \kf )$ is Hermitian, because the anti-Hermitian part $i G_2(0,\kf)$ is directly proportional to the spectral function $A(0,\kf)$ and hence vanishes, as can be seen from eq.~\eqref{gaherm}. 

\begin{figure*}
  \centering
  \begin{tabular}{cc}
  \includegraphics[width=0.49\linewidth]{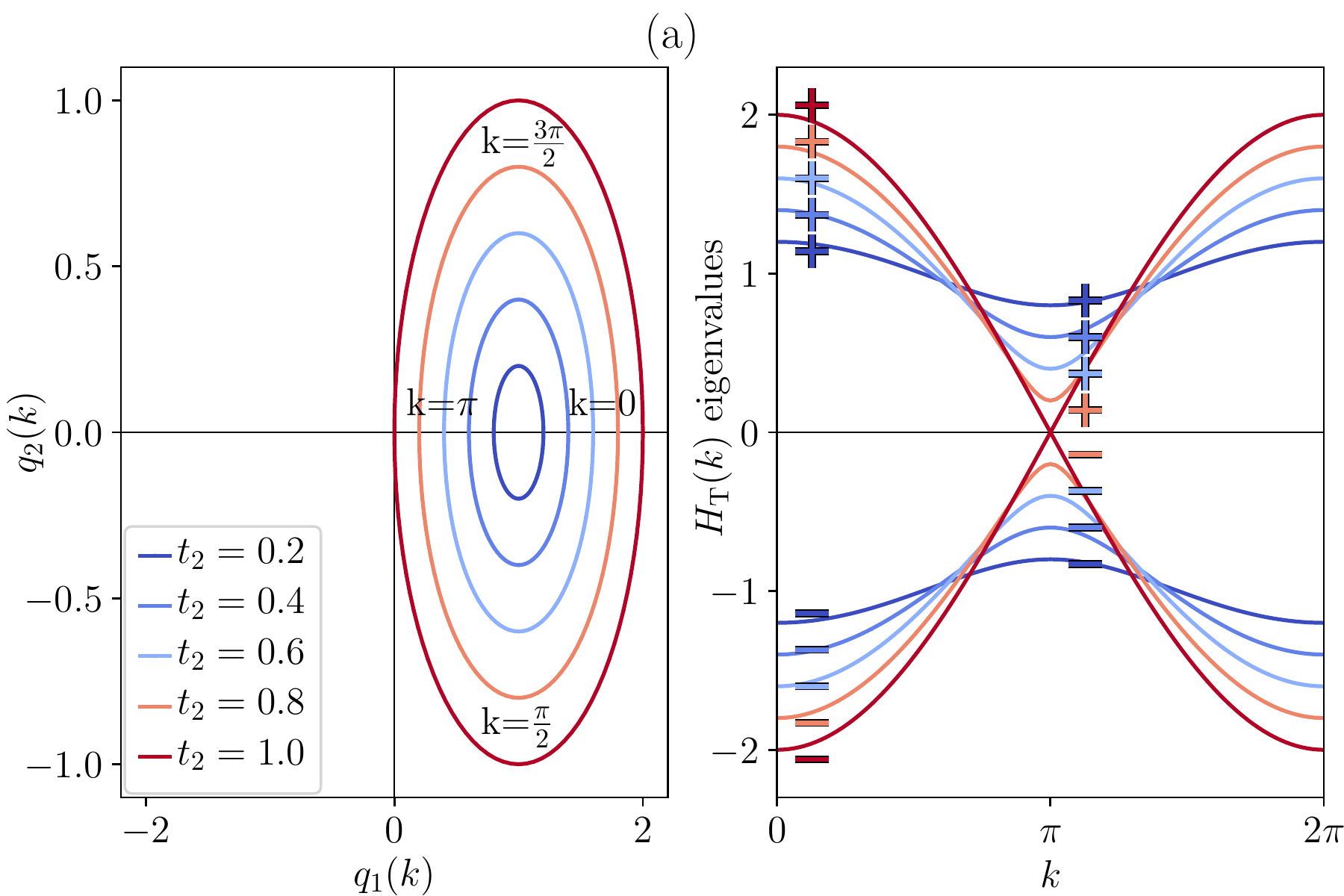} & \includegraphics[width=0.49\linewidth]{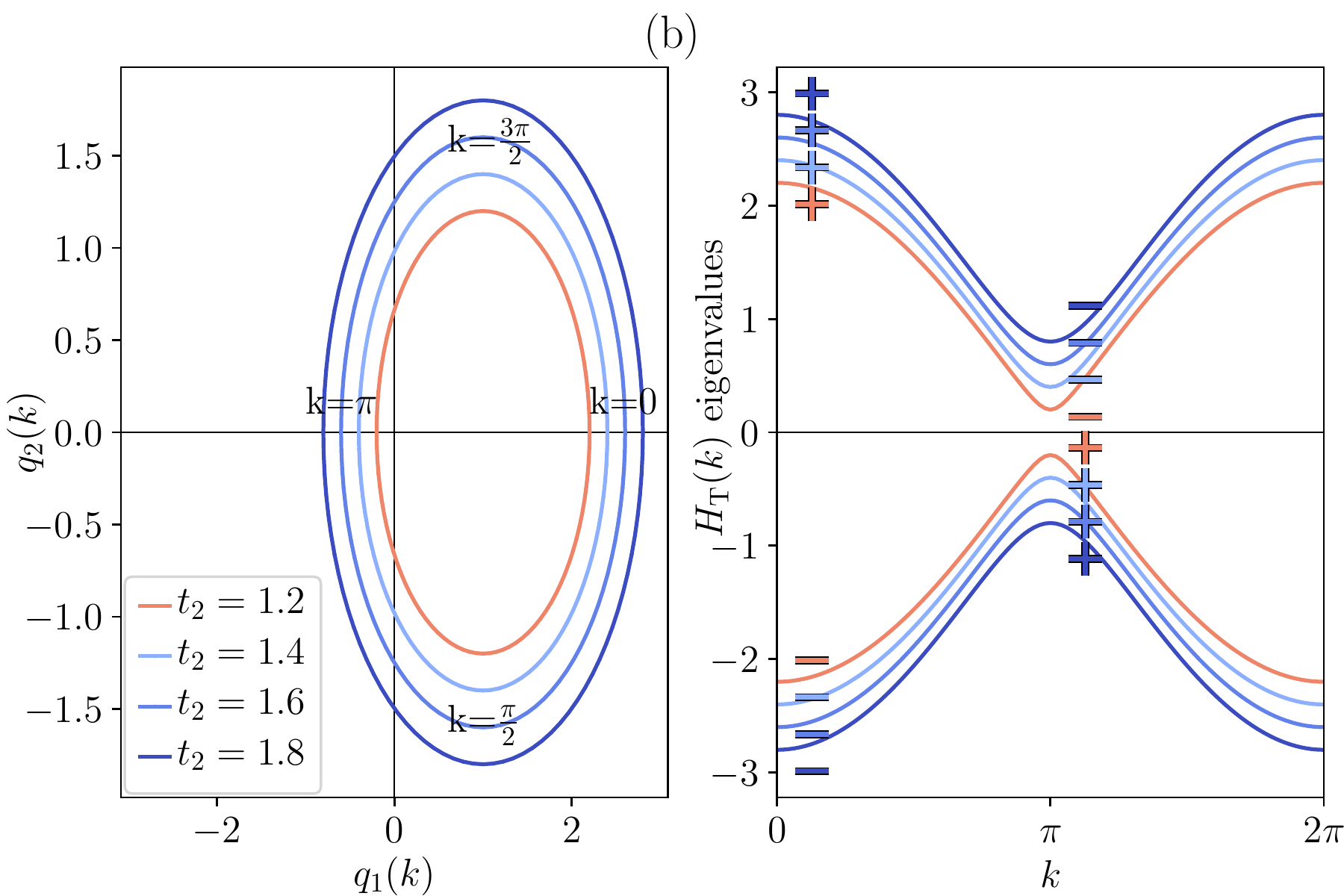} \\
  \end{tabular}
\caption{\label{noninttr} Illustration of the topological invariant $N_1$ for the (non-interacting) SSH model with $t_1 =1$ fixed. The respective left plot shows the winding of the Pauli matrix expansion coefficients $q_1(k)$ and $q_2(k)$ around the origin when $k$ sweeps the Brillouin zone (see eq.~\eqref{pauliexp}). The respective right plot shows the band structure with inversion eigenvalues of the eigenstates at the high-symmetry k-points, denoted by $+$ or $-$. The inversion eigenvalues fully determine the irreps. The value of $N_1$ can already be read of from looking at the inversion eigenvalues. In the $N_1 = 0$ phase the inversion values within the same band have the same sign, in the $N_1 = 2$ phase they have opposite signs. In Fig.~(a) for $t_1 >t_2$ one finds $N_1 = 0$ except for the point of the transition at $t_1 = t_2 =1$ where the gap closes. In Fig.~(b) for $t_2 > t_1$ one finds $N_1 = 2$.}
\end{figure*}

We now show that for a gapped spectral function the absolute values of the eigenvalues of $G(0, \kf )$ are bounded from above. To our knowledge this has not been shown before. 
Since the gap is non-zero, there exists an $\epsilon > 0$ such that the interval $\left[ - \epsilon , \epsilon \right] $ lies inside the gap. For a vector $\f{a}$ with the same dimension as $G$ and $||\f{a}|| = 1$ we can make the following estimate
\begin{align}
&|\sum_{\alpha \beta} \bar{a}_{\alpha} G_{\alpha \beta}(0 , \kf)a_{\beta}| \nonumber \\
&\leq  \sum_{\alpha \beta}\int_{- \infty}^{\infty} \frac{d \omega'}{2\pi}  \bar{a}_{\alpha} A_{\alpha \beta} (\omega', \kf) a_{\beta } \frac{1}{|\omega'| } \nonumber \\
&= \sum_{\alpha \beta} \int_{\mathbb{R}\setminus \left[ - \epsilon, \epsilon \right] }\frac{d \omega'}{2\pi} \bar{a}_{\alpha} A_{\alpha \beta} (\omega', \kf) a_{\beta }\frac{1}{| \omega' |} \nonumber \\
&\leq \frac{1}{\epsilon} \sum_{\alpha \beta}  \int_{\mathbb{R}\setminus \left[ - \epsilon, \epsilon \right] }\frac{d \omega'}{2\pi} \bar{a}_{\alpha} A_{\alpha \beta} (\omega', \kf) a_{\beta } \nonumber \\
&= \frac{1}{\epsilon}. 
\end{align}
Going from the first to the second line we have used that the spectral function $A(\omega,\kf)$ is positive semi-definite. 
Using these properties it follows that the eigenvalues of $G(0, \kf )$ are real and and their absolute value is bounded.
It follows that the topological Hamiltonian, which for $G(0, \kf )$ non-singular (i.e. $\det \big(G(0, \kf ) \big) \neq 0$) can be defined by eq.~\eqref{htopdef}, has real eigenvalues with their absolute value being bounded from below. 

\section{Relationship between Green's function topological invariants and EBRs for the SSH+U Model}
\label{app:ssh_help}

As a consequence of inversion symmetry, we have according to eq.~\eqref{hTbr}
\begin{equation}
H_{\nk{T}}(k) = \sigma_1 H_{\nk{T}}( - k)  \sigma_1 .
\label{htinvssh}
\end{equation}
For the Pauli matrix expansion coefficients in eq.~\eqref{pauliexp} this implies
\begin{align}
q_1(k)  &= q_1(-k) , \label{q1sym}\\
q_2(k)  &= -q_2(-k)  . \label{q2sym}
\end{align}
So this symmetry further restricts the form of the topological Hamiltonian.
The eigenvalues of the topological Hamiltonian are implicitly given by 
\begin{equation}
\mu_{\pm}(k) = \pm \sqrt{\big(q_1(k) \big) ^2 + \big( q_2(k) \big) ^2}
\end{equation}
For the eigenvectors of the lower and the upper band $v_{-}(k)$ and $v_{+}(k)$ we find
\begin{align}
v_{-}(k) &= \frac{1}{\sqrt{2}} \begin{pmatrix} 
 1  \\
\frac{-q(k)}{|q(k)|} \\ 
\end{pmatrix} , \\
v_{+}(k) &= \frac{1}{\sqrt{2}} \begin{pmatrix} 
 1  \\
\frac{q(k)}{|q(k)|} \\ 
\end{pmatrix}.
\label{vm}
\end{align}
Of special interest are the high symmetry k-points $\kappa = 0, \pi$, where inversion symmetry implies that $q_2(\kappa)$ must vanish, so that $q(\kappa) = q_1(\kappa )$. The inversion eigenvalues can simply be calculated by multiplying the representation of the inversion operator with the eigenvectors. For both bands we get
\begin{align}
\rho_{\mathfrak{G}}^{\kappa}(\mathcal{I}) v_{-}(\kappa) &= -\nk{sign} \big( q_1(\kappa ) \big) v_{-}(\kappa), \label{inveigm}\\
\rho_{\mathfrak{G}}^{\kappa}(\mathcal{I}) v_{+}(\kappa) &= \nk{sign} \big( q_1(\kappa ) \big) v_{+}(\kappa).
\end{align}
The evolution of the spectrum, inversion eigenvalues, and Hamiltonian parameters for the non-interacting limit $U = 0$ are shown in Fig.~\ref{noninttr}.

If the inversion eigenvalues have the same sign at both $\kappa=0$ and $\kappa=\pi$, it follows from eq.~\eqref{inveigm} that $q(0)$ and $q(\pi )$ also have the same sign.
From eq.~\eqref{q1sym} and eq.~\eqref{q2sym} we get that the $q(k)$ curve is mirror symmetric with respect to the $q_1$ axis. Now it is easy to see that the winding must be $2n$ in this case.
Suppose the contrary would be true and we would wind $2n+1$ times around the origin. Then in the interval $ \left[ 0 , \pi \right] $ $q(k)$ winds exactly an integer and a half times around the origin, because of $q_1(k) = q_1(-k)$ and $q_2(k) = -q_2(-k)$ which also implies $q_2(\pi) =0$ (imagine winding in the exact opposite direction while $k$ goes backwards form $2 \pi$ to $\pi$). But this would mean $q_1(0)$ and $q_1(\pi )$ have different signs. Hence the winding number must be $2n$.

With the same symmetry arguments it follows that if at both high symmetry k-points the inversion eigenvalues have the opposite sign and hence the signs of $q_1(0)$ and $q_1(\pi )$ are opposite, then the winding number must be $2n+1$. 
Note that if $q_2(k)=0$ for all $k$ and $q_1(0)$ and $q_1(\pi )$ have different signs then then there must be a k-point for which both $q_1$ and $q_2$ either vanish or become infinite at the same time. But this is excluded by the assumptions that we have a gap in the spectral function and $G(0,\kf)$ being non singular. So we really have to wind $2n +1$ times around the origin.

To summarize: If in the SSH+U model the inversion eigenvalues of the lower band at the high-symmetry k-points have the same sign, then $N_1 = 2n$ with $n \in \mathbb{Z}$. If they have opposite signs, then $N_1 = 2n +1$ with $n \in \mathbb{Z}$.
Spin degeneracy gives a further factor of two, i.e. $N_1 = 4n +2$ if both inversion eigenvalues have opposite sign and $N_1 = 4n $ if the signs are equal.
\bibliography{main}

\end{document}